\documentclass{aa}
\usepackage{txfonts}
\usepackage{graphicx}
\usepackage{natbib}
\bibpunct{(}{)}{;}{a}{}{,}

\begin{document}

\title{The Arches cluster revisited: II. A massive eclipsing spectroscopic
  binary in the Arches cluster}
\author{M.~E.~Lohr\inst{\ref{inst1}}\and
  J.~S.~Clark\inst{\ref{inst1}}\and F.~Najarro\inst{\ref{inst2}}\and
  L.~R.~Patrick\inst{\ref{inst3},\ref{inst4},\ref{inst5}}\and
  P.~A.~Crowther\inst{\ref{inst6}}\and
  C.~J.~Evans\inst{\ref{inst7}}}
\institute{School of Physical Sciences, The Open University, Walton
  Hall, Milton Keynes MK7\,6AA,
  UK\\ \email{Marcus.Lohr@open.ac.uk}\label{inst1}\and Centro de
  Astrobiolog\'{i}a (CSIC-INTA), Ctra. de Torrej\'{o}n a Ajalvir md-4,
  28850 Torrej\'{o}n de Ardoz, Madrid, Spain\label{inst2}\and
  Instituto de Astrof\'{i}sica de Canarias, E-38205 La Laguna,
  Tenerife, Spain\label{inst3}\and Universidad de La Laguna, Dpto.
  Astrof\'{i}sica, E-38206 La Laguna, Tenerife, Spain\label{inst4}\and
  Institute for Astronomy, University of Edinburgh, Royal Observatory
  Edinburgh, Blackford Hill, Edinburgh EH9\,3HJ, UK\label{inst5}\and
  Department of Physics and Astronomy, University of Sheffield,
  Sheffield S3\,7RH, UK\label{inst6}\and UK Astronomy Technology
  Centre, Royal Observatory, Blackford Hill, Edinburgh EH9\,3HJ,
  UK\label{inst7}}
\date{Received 19 January 2018 / Accepted 14 April 2018}

\abstract{We have carried out a spectroscopic variability survey of
  some of the most massive stars in the Arches cluster, using $K$-band
  observations obtained with SINFONI on the VLT.  One target, F2,
  exhibits substantial changes in radial velocity; in combination with
  new KMOS and archival SINFONI spectra, its primary component is
  found to undergo radial velocity variation with a period of
  10.483$\pm$0.002~d and an amplitude of
  $\sim$350~km~s\textsuperscript{-1}.  A secondary radial velocity
  curve is also marginally detectable.  We reanalyse archival
  NAOS-CONICA photometric survey data in combination with our radial
  velocity results to confirm this object as an eclipsing SB2 system,
  and the first binary identified in the Arches.  We model it as
  consisting of an 82$\pm$12~M$_{\sun}$ WN8--9h primary and a
  60$\pm$8~M$_{\sun}$ O5--6 Ia$^+$ secondary, and as having a slightly
  eccentric orbit, implying an evolutionary stage prior to strong
  binary interaction.  As one of four X-ray bright Arches sources
  previously proposed as colliding-wind massive binaries, it may be
  only the first of several binaries to be discovered in this cluster,
  presenting potential challenges to recent models for the Arches' age
  and composition.  It also appears to be one of the most massive
  binaries detected to date; the primary's calculated initial mass of
  $\gtrsim$120~M$_{\sun}$ would arguably make this the most massive
  binary known in the Galaxy.}

\titlerunning{II. A massive eclipsing spectroscopic binary in the Arches
  cluster}
 \authorrunning{M.~E.~Lohr et al.}

\maketitle

\section{Introduction}

The Arches cluster, discovered about 20 years ago
\citep{nagata1995,cotera1996} near the Galactic centre, contains a
remarkable, dense population of young, very massive stars.  Using
$K$-band spectroscopy, around 200 O and OIf$^+$ supergiants and a
dozen WNLh-type Wolf-Rayet stars have been identified
\citep{blum2001,figer2002,martins2008}, and the total cluster mass is
of the order of $10^4$~M$_{\sun}$ \citep{stolte2002}.  It is thus of
great value for investigations of the formation and evolution of the
most massive stars, and even for constraining the upper stellar mass
limit \citep{figer2005}.  It is also of great significance for our
understanding of star formation near the centre of our Galaxy, along
with the Quintuplet cluster and Galactic central cluster itself.

Identifying massive binaries in the Arches would be useful in several
ways.  Such objects are of intrinsic interest as candidate progenitors
for supernovae of various types, gamma-ray bursts and even merging
binary stellar-mass black holes like GW150914 \citep{abbott2016}.
Moreover, a double-lined spectroscopic and eclipsing binary would
permit direct mass determinations for its components, providing a
check on masses estimated from model evolutionary tracks.  This in
turn would have implications for the age of the cluster:
\citet{martins2008} agreed with \citeauthor{figer2002}'s
\citeyearpar{figer2002} estimate of 2.5$\pm$0.5~Myr, at least for the
most massive, luminous stars, but \citet{schneider2014} have argued
that an age of 3.5$\pm$0.7~Myr is more plausible, on the basis of
fitting their population synthesis models to a stellar mass function
for the cluster, with an associated conclusion that the most massive
cluster members are rejuvenated products of binary interaction and
merger.  The detection of existing massive binaries would contribute
to the resolution of this debate.

There are already indications of possible colliding-wind binaries in
the Arches from radio \citep{lang2001,lang2005} and X-ray
\citep{wang2006} observations.  A candidate massive contact eclipsing
binary was also proposed as a preliminary result of a photometric
variability survey \citep{markakis2011}.  However, the detection of
regular radial velocity (RV) variations allows binaries to be
confirmed and, in combination with suitable light curves, their masses
constrained (e.g. \citealp{ritchie2009,ritchie2010,clark2011}).
Therefore we have undertaken a multi-epoch spectroscopic survey of the
most massive members of the Arches, reported in \citet{clark2018} and
\citet{lohr2018}, hereafter Papers I and III.  Here, we describe our
detection and investigation of the target which gave the strongest
evidence for binarity, and for which sufficient additional data were
available to permit preliminary modelling.

\section{Data acquisition and reduction}

\subsection{Spectroscopy}

This section focuses upon the observations and data reduction specific
to the binary which is the subject of this paper.  For full details of
the reduction procedures used and the nature of the wider
spectroscopic survey, see Paper I.

The SINFONI integral field spectrograph on the ESO/VLT
\citep{eisenhauer2003,bonnet2004} was used in service mode to observe
fields in the central Arches cluster, and on the periphery of the
cluster, in the $K$ band, during April to August 2011 and March to
August 2013\footnote{ESO proposals 087.D-0317 and 091.D-0187, PI
  J.~S.~Clark.}.  The data cubes were reduced, and the stellar spectra
extracted, as described in Paper I.  Objects were observed on up to
seven distinct epochs; the main subject of this paper, F2 (i.e. the
second object on the list of \citet{figer2002}) was observed on five
epochs.  When selecting the pixels of F2 for extraction from the data
cubes, care had to be taken to avoid pixels contaminated by light from
a nearby fainter cluster member (F19).  A further epoch of SINFONI
spectra from 2005, for outlying fields which included F2, was
similarly extracted from data cubes used for
\citet{martins2008}\footnote{ESO proposal 075.D-0736, PI T.~Paumard.}.
These cubes were provided by F.~Martins, with sky subtraction and
telluric removal already carried out as described in that paper.

\begin{table}
\caption{Spectroscopic observation log for F2.  S/N ratios were
  measured for three central wavelength regions free of significant
  spectral lines.}
\label{spectable}
\centering
\begin{tabular}{c c c c}
\hline\hline\noalign{\smallskip}
Obs. date & BJD(TDB) & Instrument & S/N \\
(YYMMDD) & $-$2450000 &  &  \\
\hline
990704 & 1363.945 & Keck/NIRSPEC & 34 \\
050610 & 3531.755 & VLT/SINFONI\tablefootmark{b} & 243 \\
110414 & 5665.819 & VLT/SINFONI\tablefootmark{c} & 184 \\
110419 & 5670.716 & VLT/SINFONI\tablefootmark{c} & 142 \\
110425 & 5676.808 & VLT/SINFONI\tablefootmark{c} & 168 \\
110504 & 5685.820 & VLT/SINFONI\tablefootmark{c} & 231 \\
110521 & 5702.648 & VLT/SINFONI\tablefootmark{c} & 195 \\
110623 & 5735.816 & VLT/SINFONI\tablefootmark{c} & 179 \\
110626 & 5738.726 & VLT/SINFONI\tablefootmark{c} & 217 \\
110629\tablefootmark{a} & 5741.735 & VLT/SINFONI\tablefootmark{c,d} & 304 \\
110630 & 5742.683 & VLT/SINFONI\tablefootmark{c} & 227 \\
110824 & 5797.561 & VLT/SINFONI\tablefootmark{c} & 179 \\
110827 & 5800.544 & VLT/SINFONI\tablefootmark{d} & 137 \\
110828 & 5801.590 & VLT/SINFONI\tablefootmark{d} & 60 \\
110829 & 5802.550 & VLT/SINFONI\tablefootmark{c} & 166 \\
111002 & 5836.512 & VLT/SINFONI\tablefootmark{c} & 174 \\
130717 & 6490.715 & VLT/SINFONI\tablefootmark{d} & 172 \\
130808 & 6512.611 & VLT/SINFONI\tablefootmark{d} & 142 \\
140430 & 6777.831 & VLT/KMOS & 72 \\
140723 & 6861.523 & VLT/KMOS & 69 \\
140804 & 6873.737 & VLT/KMOS & 63 \\
140805 & 6874.692 & VLT/KMOS & 90 \\
140811\tablefootmark{a} & 6880.596 & VLT/KMOS & 75 \\
140812 & 6881.622 & VLT/KMOS & 72 \\
140813 & 6882.631 & VLT/KMOS & 76 \\
\hline
\end{tabular}
\tablefoot{
\tablefoottext{a}{Epoch combined from two observations made within a few hours of each other.}
\tablefoottext{b}{From data cube supplied by F.~Martins.}
\tablefoottext{c}{Archival data: ESO proposal 087.D-0342, PI
  G.~Pietrzynski.}
\tablefoottext{d}{New survey data (Paper II).}
}
\end{table}

Our search for RV variability in the brighter cluster
members (described fully in Paper 3) indicated immediately that F2
stood out as exhibiting highly significant
($\sigma_\mathrm{detect}$=33.75) and very substantial
($\Delta$RV=361$\pm$11~km~s\textsuperscript{-1}) variability across
the six epochs available.  Other spectra for this object were
therefore sought.  Raw data for 12 more SINFONI spectra of F2 from
2011 were extracted from the ESO archive\footnote{ESO proposal
  087.D-0342, PI G.~Pietrzynski.} and reduced as described above.  The
Keck/NIRSPEC spectrum from 1999, used to classify F2 in
\citet{figer2002}, was supplied, reduced as described in that paper.
Eight $K$-band spectra from a 2014 kinematic survey of the Galactic
centre using KMOS \citep{sharples2013} on the VLT\footnote{ESO
  proposal 093.D-0306, PI J.~S.~Clark.} were also provided.  These
were reduced using the KMOS pipeline \citep{davies2013}, with routines
for sky and telluric correction adapted from those described by
\citet{patrick2015}, with an additional stage of dividing the science
spectra into three sections prior to cross-correlation with telluric
spectra, to optimize the telluric removal.

\begin{figure*}
\centering
\includegraphics[width=17cm]{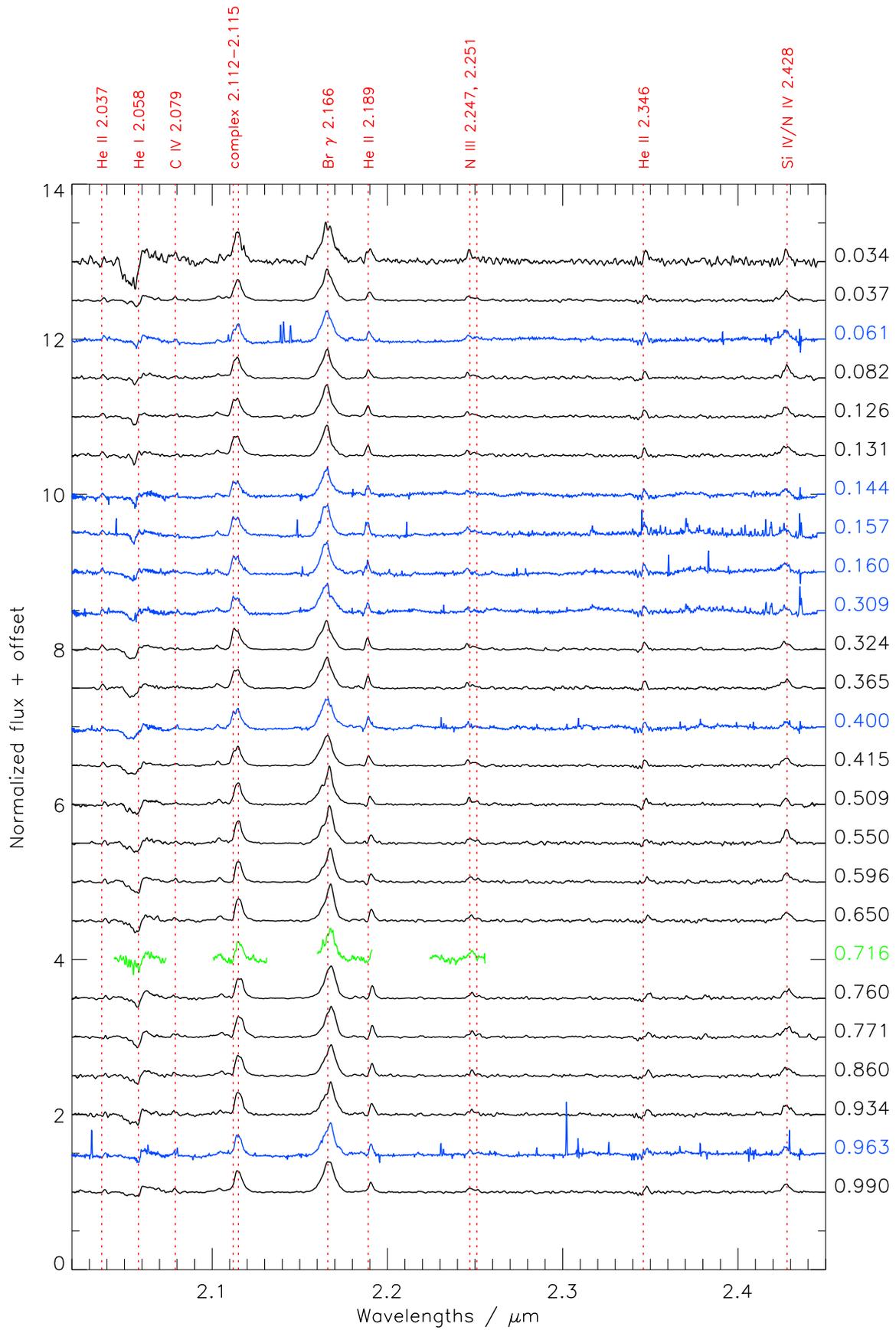}
\caption{Normalized barycentric-corrected spectra for F2, ordered by
  phase (shown on right), and with a constant offset in the
  $y$-direction.  Dashed vertical lines (red in online version)
  indicate rest wavelengths of significant lines.  Spectra from
  SINFONI reduced by the same method are shown in black; those from
  KMOS are in blue and the spectrum from NIRSPEC is in green.}
\label{specarray}
\end{figure*}

SINFONI provides a resolving power (at our plate scales of 0\farcs1 or
0\farcs025) of $R\sim4500$ at 2.2~$\mu$m, and KMOS a little lower, at
$R\sim4250$.  All spectra were rebinned to a common dispersion of
0.000245~$\mu$m pixel\textsuperscript{-1} and common wavelength range
of 2.02--2.45~$\mu$m (except the NIRSPEC spectrum, which covered four
short wavelength regions within the $K$-band), and had barycentric
velocity corrections made where these had not already been performed.
After combination of observations from the same epoch, this left us
with 25 spectra of F2.  Table \ref{spectable} summarises the
spectroscopic observations used, and Fig.~\ref{specarray} shows all
spectra ordered by phase.

RVs were measured for F2 by cross-correlation in IRAF, using spectra
near assumed eclipses as trial templates, and using several different
lines to confirm results.  Meaningful timings for each spectrum were
determined as the average of the contributing original science frames'
mid-exposure times, converted to Barycentric Julian Dates in
Barycentric Dynamical Time
(BJD(TDB))\footnote{http://astroutils.astronomy.ohio-state.edu/time/
  (see also \citealp{eastman2010}).}.  The spectroscopic period was
found by a form of string length minimization, both singly and in
combination with photometric data (e.g. \citealp{dworetsky1983}); it
was checked using the Lomb-Scargle periodogram method
\citep{lomb1976,scargle1982,horne1986}.

\subsection{Photometry}

$K_s$ band time-series photometry of the Arches cluster had been
obtained by Pietrzynski et al.\footnote{ESO proposal 081.D-0480, PI
  G.~Pietrzynski.}  between June 2008 and April 2009 using NAOS-CONICA
(NaCo) on the VLT \citep{lenzen2003,rousset2003}.  This study was
briefly described in \citet{markakis2011,markakis2012};
the data set included all the stars modelled in \citet{martins2008}
with the exception of B1, which fell outside the field of view.
Observations covered 31 nights, and the majority consisted of 15
jittered frames, interspersed with sky frames.

We extracted the raw data from the ESO archive and carried out basic
reductions (dark and flat-field corrections, background removal and
co-adding of jittered frames) using the ESO NaCo pipeline running
under Gasgano.  39 distinct combined images were obtained (a few
short, incomplete runs containing small numbers of frames were either
folded in with adjacent runs to improve the signal or discarded as
unusable; this presumably explains the discrepancy with the 46
distinct observations reported in \citealp{markakis2011}).

\citeauthor{markakis2011} \citeyearpar{markakis2011,markakis2012}
reported serious problems with the extraction of reliable magnitudes
from this photometric data set, owing to the highly-variable PSF over
each frame associated with imperfect atmospheric correction by the
adaptive optics.  We also found that PSF modelling in IRAF gave
negligible improvement over simple aperture photometry.  Therefore
iterative PSF-modelling photometry was used on a single high
signal-to-noise image to detect blends and to determine the centroid
coordinates of all measurable sources; these coordinates were then
used as the basis for aperture photometry on all images after careful
alignment.

A custom IDL code was written to generate light curves for each
distinct source.  Since we did not know in advance which stars in the
field might be stable enough to use as reference stars for F2, those
light curves bright enough to be detected in every image were selected
and shifted to a common minimum magnitude (this produced less scatter
than using their mean magnitudes).  A clear night-to-night trend was
apparent for the majority of sources, while some (including F2) were
highly deviant.  Thirteen sources which tracked the mean trend very
closely were thus selected and combined to form a composite reference
star, allowing construction of a differential light curve for F2.

Timings were determined and converted to BJD(TDB) in a similar manner
to that used for the spectroscopic measurements.  The photometric
period was determined by a form of string length minimization, both
singly and in combination with spectroscopic data, and checked using
the Lomb-Scargle periodogram method (as for the spectroscopic period).

In addition to time-series photometry in the $K$-band, for spectral
energy distribution (SED) fitting we have used broadband NICMOS
photometric measurements of F2 from \citet{figer2002} (F110W, F160W
and F205W filters) and \citet{dong2011} (NIC3 F190N), and new WFC3
narrowband photometry (F127M, F139M and F153M) from \citet{dong2018},
reduced as described in Paper I.  In each case small corrections were
made to the magnitudes to account for the object's photometric
variability.  Uncertainties on magnitudes were provided by Dong
(personal communication) for WFC3 and F190N, and for NICMOS were
inferred from the plots in \citet{figer2002}.

\section{Results}

F2 stood out clearly amongst the 34 bright sources in our survey for
which RVs could be measured.  Fig.~\ref{specarray} shows
the spectra ordered by phase; the motion of the emission lines is
readily apparent, especially near quadrature.  The main emission lines
are Br$\gamma$, \ion{He}{I} 2.058~$\mu$m (strong P~Cygni profile),
\ion{He}{II} 2.037, 2.189, 2.346~$\mu$m (weaker P~Cygni profiles),
\ion{C}{IV} 2.079~$\mu$m, \ion{N}{III} 2.247, 2.251~$\mu$m,
\ion{Si}{IV} 2.428~$\mu$m and the blended \ion{He}{I}, \ion{N}{III},
\ion{C}{III}, \ion{O}{III} complex at 2.112-2.115~$\mu$m.  On the
basis of these, \citet{martins2008} classified F2 as WN8--9h.  We
should also note that, as discussed in Paper I, an \ion{N}{IV} line
contributes to the feature at 2.428~$\mu$m.

\begin{figure}
\resizebox{\hsize}{!}{\includegraphics{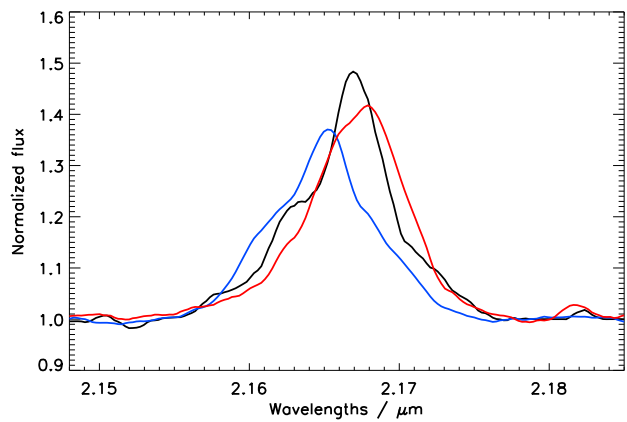}}
\caption{Br$\gamma$ line at three phases, from SINFONI spectra: in
  black, near secondary eclipse (phase 0.550); in blue and red, near
  quadratures (phases 0.324 and 0.760).}
\label{brgamma}
\end{figure}

\begin{figure}
\resizebox{\hsize}{!}{\includegraphics{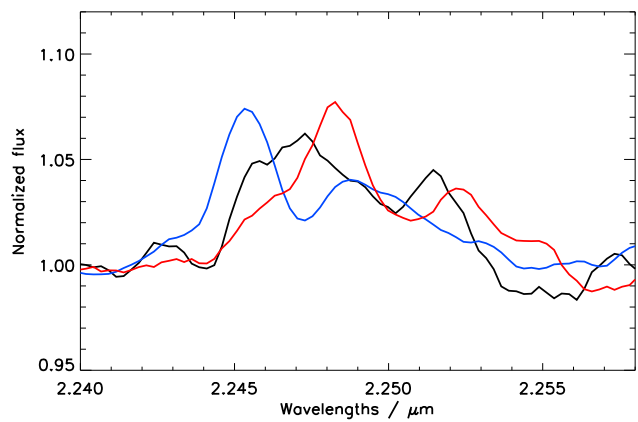}}
\caption{\ion{N}{III} 2.247, 2.251~$\mu$m lines at three phases: in
  black, near secondary eclipse (phase 0.550); in blue and red, near
  quadratures (phases 0.324 and 0.760).}
\label{NIII}
\end{figure}

There is no clear evidence for line splitting, only for shifting of
lines attributable to the assumed primary Wolf-Rayet component (with
one exception).  There are, however, some changes in the profiles of
certain lines at different phases; for instance, Br$\gamma$
(Fig.~\ref{brgamma}) shows the strongest and narrowest peak near
eclipse of the secondary component, but is slightly weakened and
broadened near quadratures, perhaps indicating a secondary emission
component in this line.  The \ion{N}{III} 2.247, 2.251~$\mu$m pair of
lines, not expected to be present in an OI star, do not show obvious
profile changes with phase; they are perhaps the purest representative
of the motion of the WNL component (Fig.~\ref{NIII}).

\begin{figure}
\resizebox{\hsize}{!}{\includegraphics{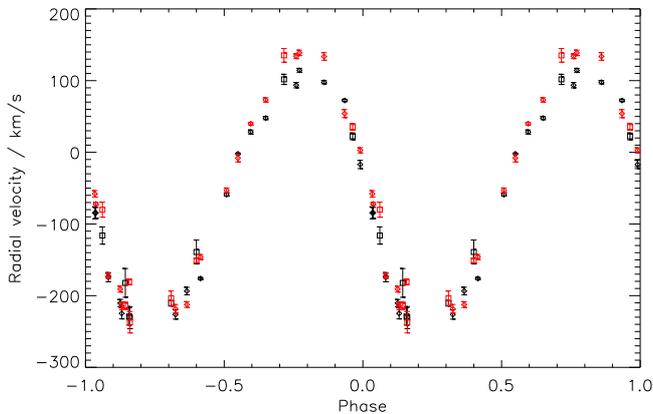}}
\caption{RV curve for assumed WNL primary of F2, measured from
  Br$\gamma$ (black) and \ion{N}{III} 2.247, 2.251~$\mu$m (red) lines
  with F1 spectrum.  SINFONI data are plotted as diamonds; KMOS and
  NIRSPEC as squares.}
\label{rv1bothplot}
\end{figure}

\begin{figure}
\resizebox{\hsize}{!}{\includegraphics{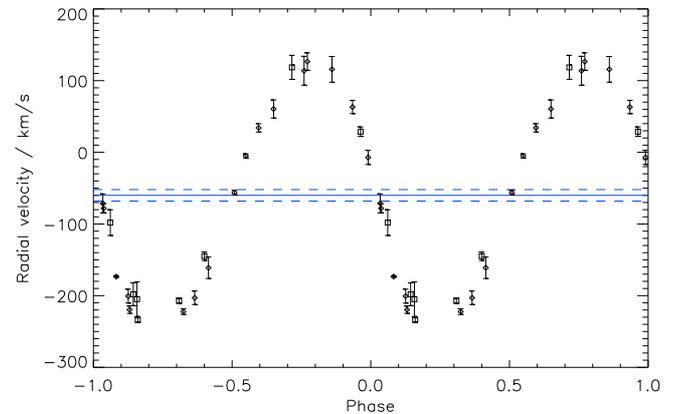}}
\caption{Combined RV curve for assumed WNL primary of F2.  SINFONI
  data are plotted as diamonds; KMOS and NIRSPEC as squares.  The
  solid blue horizontal line indicates the approximate systemic
  velocity (relative to F1); the dashed blue lines indicate the
  systemic velocities associated with the curves derived from the
  Br$\gamma$ and \ion{N}{III} lines separately, and may be taken as
  the uncertainty on the systemic velocity for the combined RV curve.}
\label{rv1combplot}
\end{figure}

\begin{table*}
\caption{RVs for F2 relative to F1.  The uncertainties
  in the final column were multiplied by two for further analysis.}
\label{rvtable}
\centering
\begin{tabular}{r r r r r r r r r r r}
\hline\hline \noalign{\smallskip} Obs. date & BJD(TDB) & Phase & RV$_1$
(Br$\gamma$) & Unc. & RV$_1$ (\ion{N}{III}) & Unc. & RV$_1$
(comb.) & Unc. & RV$_2$ (\ion{C}{IV}) & Unc. \\
(yymmdd) & $-$2450000 & & km~s\textsuperscript{-1} &
km~s\textsuperscript{-1} & km~s\textsuperscript{-1} &
km~s\textsuperscript{-1} & km~s\textsuperscript{-1} &
km~s\textsuperscript{-1} & km~s\textsuperscript{-1} &
km~s\textsuperscript{-1} \\
\hline
990704 & 1363.945 & 0.716 & 101.7 & 7.1 & 135.3 & 9.6 & 118.5 & 16.8 & & \\
050610 & 3531.755 & 0.509 & $-$59.0 & 2.1 & $-$52.9 & 3.2 & $-$56.0 & 3.0 & & \\
110414 & 5665.819 & 0.082 & $-$174.2 & 6.4 & $-$172.3 & 3.4 & $-$173.3 & 0.9 & 3.7 & 6.4 \\
110419 & 5670.716 & 0.550 & $-$2.1 & 0.9 & $-$7.9 & 5.5 & $-$5.0 & 2.9 & & \\
110425 & 5676.808 & 0.131 & $-$224.7 & 7.7 & $-$214.4 & 5.2 & $-$219.6 & 5.2 & 134.8 & 7.2 \\
110504 & 5685.820 & 0.990 & $-$17.0 & 5.8 & 2.7 & 3.7 & $-$7.1 & 9.9 & $-$65.8 & 5.4 \\
110521 & 5702.648 & 0.596 & 28.3 & 2.8 & 40.0 & 1.9 & 34.1 & 5.8 & & \\
110623 & 5735.816 & 0.760 & 93.7 & 3.7 & 133.9 & 3.2 & 113.8 & 20.1 & & \\
110626 & 5738.726 & 0.037 & $-$84.5 & 7.6 & $-$71.8 & 2.0 & $-$78.2 & 6.3 & $-$22.2 & 3.5 \\
110629 & 5741.735 & 0.324 & $-$226.3 & 6.4 & $-$218.3 & 6.0 & $-$222.3 & 4.0 & 167.5 & 8.1 \\
110630 & 5742.683 & 0.415 & $-$176.0 & 1.5 & $-$146.1 & 3.5 & $-$161.1 & 15.0 & 41.7 & 3.7 \\
110824 & 5797.561 & 0.650 & 47.8 & 2.2 & 73.1 & 3.1 & 60.4 & 12.6 & & \\
110827 & 5800.544 & 0.934 & 72.4 & 1.7 & 53.9 & 5.7 & 63.2 & 9.2 & & \\
110828 & 5801.590 & 0.034 & $-$84.5 & 8.4 & $-$58.0 & 4.5 & $-$71.2 & 13.3 & $-$28.1 & 5.1 \\
110829 & 5802.550 & 0.126 & $-$210.3 & 5.1 & $-$190.9 & 4.1 & $-$200.6 & 9.7 & 68.4 & 8.5 \\
111002 & 5836.512 & 0.365 & $-$193.5 & 5.3 & $-$212.3 & 4.0 & $-$202.9 & 9.4 & 121.8 & 5.6 \\
130717 & 6490.715 & 0.771 & 114.4 & 2.7 & 138.8 & 3.8 & 126.6 & 12.2 & & \\
130808 & 6512.611 & 0.860 & 97.8 & 2.3 & 133.7 & 5.3 & 115.8 & 18.0 & & \\
140430 & 6777.831 & 0.160 & $-$230.6 & 15.4 & $-$237.1 & 14.9 & $-$233.8 & 3.3 & & \\
140723 & 6861.523 & 0.144 & $-$182.2 & 20.1 & $-$213.9 & 5.1 & $-$198.0 & 15.9 & & \\
140804 & 6873.737 & 0.309 & $-$210.6 & 4.0 & $-$203.6 & 10.1 & $-$207.1 & 3.5 & 168.5 & 14.4 \\
140805 & 6874.692 & 0.400 & $-$139.0 & 16.7 & $-$150.9 & 3.1 & $-$145.0 & 6.0 & & \\
140811 & 6880.596 & 0.963 & 22.2 & 4.7 & 35.4 & 4.7 & 28.8 & 6.6 & & \\
140812 & 6881.622 & 0.061 & $-$116.1 & 12.0 & $-$80.1 & 10.5 & $-$98.1 & 18.0 & & \\
140813 & 6882.631 & 0.157 & $-$229.0 & 12.1 & $-$180.8 & 4.1 & $-$204.9 & 24.1 & 144.2 & 6.3 \\
\hline
\end{tabular}
\end{table*}

RVs were measured for the assumed primary WNL component from these two
diagnostic features (Br$\gamma$ and the \ion{N}{III} pair), since they
are strong, clearly visible in all spectra, in a wavelength region
uncontaminated by strong residual telluric features, and expected to
be relatively unaffected by a less massive companion.  The final
combined spectrum of F1 (see Paper I) was used as a template, since it
is also a WNL star with a very similar spectral appearance to F2, and
was not found to exhibit significant radial velocity variability.
Fig.~\ref{rv1bothplot} shows the results; a full amplitude of
variability around 350~km~s\textsuperscript{-1} is clearly apparent
from both features, far exceeding the uncertainties on the
measurements.  The NIRSPEC and KMOS spectra fit the trend of the
SINFONI data quite well, despite having rather larger uncertainties.
The curve from the \ion{N}{III} lines is arguably of slightly greater
amplitude than that from Br$\gamma$, but is also somewhat more
scattered, owing to its lower S/N.  Therefore, we use the average of
the RVs measured from the two features for further analysis, and take
the two original velocities as the uncertainty bounds for each new
value (see Fig.~\ref{rv1combplot} and Table~\ref{rvtable}).  We may
note here a surprisingly large systemic velocity offset between F2 and
F1 (about 60~km~s\textsuperscript{-1}); a preliminary determination of
systemic velocities for all bright cluster members, to be finalised in
Paper III, suggests that F2 is the outlier.

\begin{figure}
\resizebox{\hsize}{!}{\includegraphics{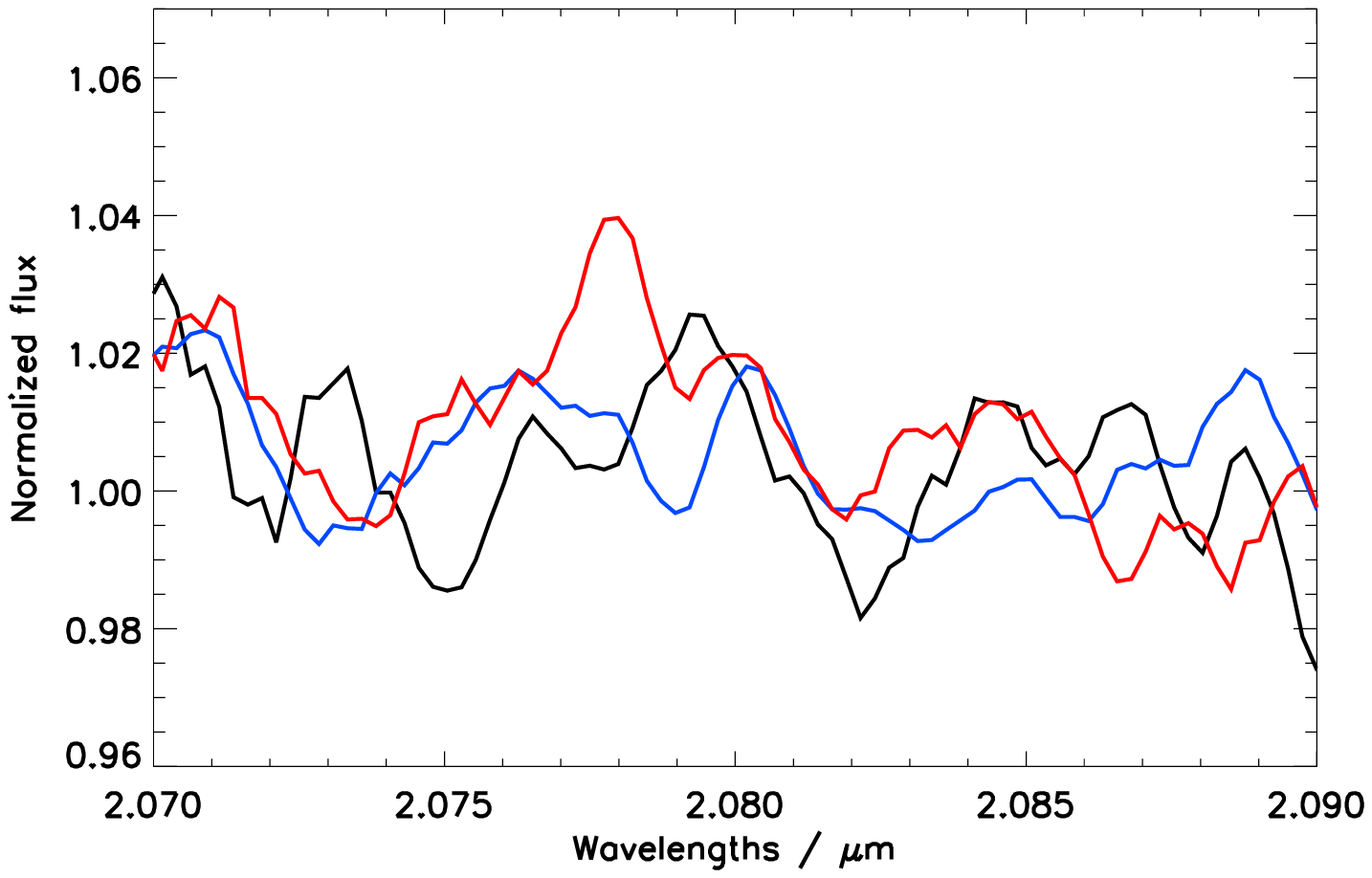}}
\caption{\ion{C}{IV} 2.079~$\mu$m line at three phases: in black, near
  secondary eclipse (phase 0.550); in blue and red, near quadratures
  (phases 0.324 and 0.760).  Although the line is faint and noisy, we
  may observe that the main peak moves in anti-phase with the other
  lines (e.g. Figs.~\ref{brgamma} and \ref{NIII}).}
\label{CIV}
\end{figure}

\begin{figure}
\resizebox{\hsize}{!}{\includegraphics{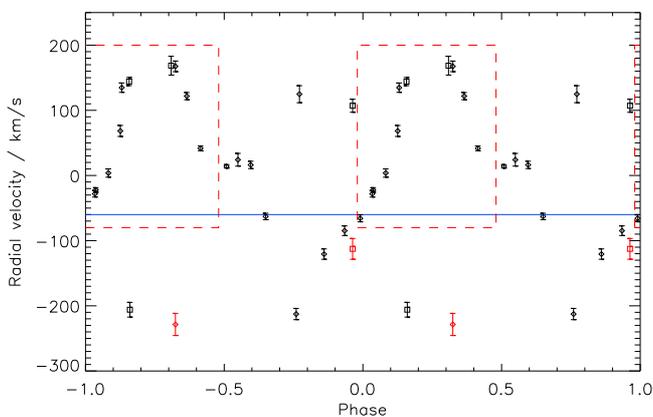}}
\caption{RV curve measured from \ion{C}{IV} 2.079~$\mu$m line.
  SINFONI data are plotted as diamonds and KMOS as squares.  The
  horizontal line indicates the approximate systemic velocity
  (relative to F1) found from the primary curve
  (Fig.~\ref{rv1combplot}).  Most of the data points indicate the
  velocity corresponding to a single-peaked cross-correlation
  function, or the stronger peak of a deblended double-peaked
  function; the two red data points indicate a weaker but still
  significant peak in a deblended double-peaked function.  The
  velocities taken as indicative of the behaviour of the secondary
  component are surrounded by the red dashed line, and are included in
  Table~\ref{rvtable}.}
\label{rv2plot}
\end{figure}

It is arguable whether the spectra show clear evidence of secondary
motion.  The most promising feature is the \ion{C}{IV} line at
2.079~$\mu$m, which appears weakly in emission in most spectra, and
seems to vary in anti-phase with the other emission lines where
present (Fig.~\ref{CIV}).  This is expected to be one of the strongest
emission lines of an O supergiant (see Paper I), though it is also a
weak emission feature in some of the Arches WNL spectra, including
that of F1.  An attempt was therefore made to measure the radial
velocities of this line by cross-correlation, again using the combined
spectrum of F1 as a template, and the results are shown in
Table~\ref{rvtable} and Fig.~\ref{rv2plot}.

Anti-phase motion with amplitude greater than the primary is clearest
from around phase 0.0 to 0.4 i.e. when the primary is partially
eclipsed by the secondary and shortly afterwards.  One KMOS point at
phase 0.160 deviates from the general trend; however, this data point
and the one at phase 0.324 (corresponding to the weaker peak deblended
from a double-peaked cross-correlation function) both match the
expected velocities of the primary WNL star at these phases
(Fig.~\ref{rv1combplot}), and so may be assumed to result from that
component.  This correspondence of primary RVs measured near
quadrature from the Br$\gamma$, \ion{N}{III} and \ion{C}{IV} lines
supports our adoption of the same systemic velocity for the secondary
RV curve, as shown in Fig.~\ref{rv2plot}.

Around phases 0.5--0.6, the velocities are again close to
those expected for the primary, which is unsurprising given that the
secondary is partially eclipsed here.  Between phases 0.6 and 0.0 the
velocities are more scattered, with those at phases 0.771 and 0.963
being close to the expected values for the primary, and other data
points exhibiting a partial blueshift, though with amplitude smaller
than that of the primary.  It seems implausible that these points
reflect the true motion of the secondary; they may instead correspond
to an average of the primary and secondary velocities in a
wind-collision region of the binary.  By contrast, the data points
from phases 0.0 to 0.4 can at worst only be underestimates of the true
secondary velocities; it is hard to see a mechanism whereby they could
systematically overestimate the amplitude of the secondary's radial
velocity curve.  The asymmetry of the velocities at the two
quadratures may be caused by the primary wind wrapping round the
secondary star; additional complications in measuring the blueshifted
components of the \ion{C}{IV} line at 2.079~$\mu$m are the red wing of
the \ion{He}{I} 2.058~$\mu$m line's broad P~Cygni profile, a possible
contribution from a faint \ion{C}{IV} line at 2.070~$\mu$m (clearly
visible in WNL stars F6, F9, F14 and F16, and in O supergiants: see
Paper I), and imperfectly-corrected telluric features (a deep telluric
absorption band extends from around 2.04--2.08~$\mu$m).

Therefore we tentatively identify a subset of the RVs
measured from the \ion{C}{IV} line at 2.079~$\mu$m as corresponding to
the secondary component (Table~\ref{rvtable}), while acknowledging
that these are likely to represent minimum values.  These data points
suggest a full amplitude of variability around
500~km~s\textsuperscript{-1}.  Given the scatter in primary velocities
measured from two strong lines, the true uncertainties in our
secondary velocities may be expected to be greater than those found
through cross-correlation using a single weak line; we therefore
double the size of these uncertainties for the remainder of the
analysis, to make them comparable to the primary velocities.

\begin{figure}
\resizebox{\hsize}{!}{\includegraphics{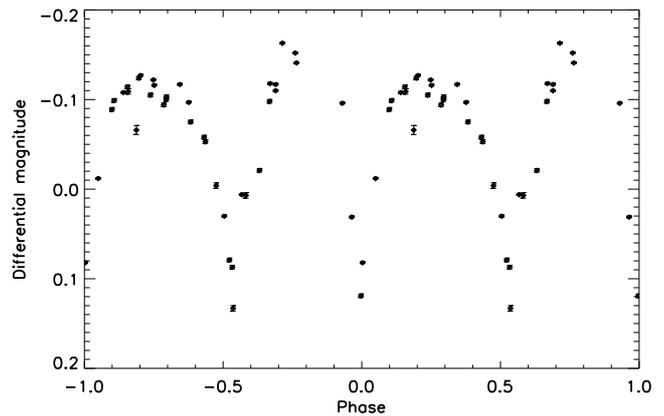}}
\caption{Light curve for F2.  The outlying point with larger
  uncertainty at phase 0.187 was associated with an incomplete
  observing run and has been excluded from further analysis.}
\label{lcplot}
\end{figure}

The light curve of F2 (Fig.~\ref{lcplot}) also shows substantial
variability, approaching 0.3 magnitudes (Table~\ref{lctable}), where
the typical scatter of the stars contributing to our composite
reference star was just 0.015 magnitudes.  One obviously deviant point
in the light curve was excluded from further analysis; the error bars
on other data points are probably too small, and the scatter of points
about the average trend would give a more realistic estimate of the
true uncertainties.  We therefore multiply all uncertainties by four
for further analysis.  It is nonetheless clear that the light curve
has the appearance of a contact or near-contact binary, with narrow
minima and continuous out-of-eclipse variability (see
e.g. \citet{lucy1968}).  Despite this, the two minima are not
separated by exactly half a cycle, suggesting a small eccentricity in
the orbit, surprising in a contact system; this may be further
supported by different heights of the two maxima, though the limited
coverage of the second maximum makes this uncertain.  The (small)
apparent difference in depths of the two minima is probably not
significant but a result of the partial coverage of these brief
regions of the orbital cycle.

\begin{table}
\caption{Differential photometry and phases for F2.  The uncertainties
in the final column were multiplied by four for further analysis.}
\label{lctable}
\centering
\begin{tabular}{c c c c}
\hline\hline\noalign{\smallskip}
BJD(TDB) & Phase &  $K$ band & Uncertainty \\
$-$2450000 &  & (diff. mag.) & (mag.) \\
\hline
4623.890 & 0.690 & $-$0.110 & 0.001 \\
4627.664 & 0.050 & $-$0.012 & 0.001 \\
4639.753 & 0.203 & $-$0.127 & 0.001 \\
4647.730 & 0.964 & 0.031 & 0.001 \\
4651.714 & 0.344 & $-$0.117 & 0.001 \\
4666.558 & 0.760 & $-$0.152 & 0.001 \\
4670.698 & 0.155 & $-$0.114 & 0.002 \\
4671.675 & 0.248 & $-$0.122 & 0.001 \\
4671.709 & 0.252 & $-$0.116 & 0.001 \\
4676.555 & 0.714 & $-$0.163 & 0.001 \\
4679.523 & 0.997 & 0.119 & 0.002 \\
4682.550 & 0.286 & $-$0.094 & 0.002 \\
4683.497 & 0.376 & $-$0.097 & 0.001 \\
4685.497 & 0.567 & 0.006 & 0.001 \\
4686.567 & 0.669 & $-$0.118 & 0.001 \\
4687.568 & 0.765 & $-$0.141 & 0.001 \\
4691.510 & 0.141 & $-$0.108 & 0.001 \\
4692.528 & 0.238 & $-$0.105 & 0.002 \\
4694.558 & 0.431 & $-$0.058 & 0.002 \\
4694.606 & 0.436 & $-$0.053 & 0.002 \\
4695.515 & 0.523 & 0.079 & 0.002 \\
4700.547 & 0.003 & 0.082 & 0.001 \\
4701.557 & 0.099 & $-$0.089 & 0.002 \\
4701.642 & 0.107 & $-$0.099 & 0.002 \\
4702.481 & 0.187 & $-$0.066 & 0.005 \\
4702.575 & 0.196 & $-$0.124 & 0.002 \\
4703.611 & 0.295 & $-$0.103 & 0.002 \\
4704.527 & 0.382 & $-$0.075 & 0.002 \\
4705.497 & 0.475 & $-$0.004 & 0.003 \\
4707.515 & 0.667 & $-$0.098 & 0.002 \\
4712.656 & 0.158 & $-$0.109 & 0.003 \\
4716.588 & 0.533 & 0.087 & 0.002 \\
4716.616 & 0.536 & 0.133 & 0.003 \\
4717.617 & 0.631 & $-$0.021 & 0.002 \\
4724.572 & 0.294 & $-$0.100 & 0.002 \\
4727.583 & 0.582 & 0.007 & 0.003 \\
4919.900 & 0.927 & $-$0.096 & 0.001 \\
4925.921 & 0.502 & 0.030 & 0.001 \\
4927.863 & 0.687 & $-$0.117 & 0.001 \\
\hline
\end{tabular}
\end{table}

\begin{figure}
\resizebox{\hsize}{!}{\includegraphics{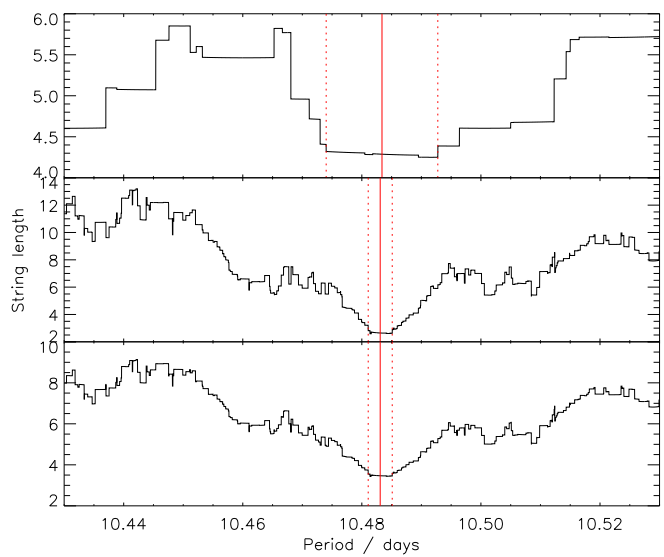}}
\caption{String length values for candidate F2 periods near expected
  period: for the light curve data (upper panel), WNL radial
  velocities (middle panel) and combined light curve and radial
  velocities (lower panel).  Dotted red vertical lines indicate the
  bounds of the region in each panel where the string length statistic
  is relatively constant and near its minimum value.  Solid red
  vertical lines indicate the mid-point of these ranges.}
\label{stringlength}
\end{figure}

Period searches for the light curve by string length minimization
indicate a minimum around 10.49~d (as found by
\citealp{markakis2011}), but with very similar values of the
minimization statistic between 10.474 and 10.493~d, giving a mid-point
of 10.483$\pm$0.009~d.  Using the primary RV curve alone,
a much better-constrained minimum is found within the same range,
between 10.481 and 10.485~d. A joint determination of the best period
for both light and RV curves using string length
minimization (after normalizing each curve by its amplitude) gives the
same narrow minimum of 10.483$\pm$0.002~d: the RVs,
though less numerous than the photometric data points, cover a far
longer time base, and so provide a stronger constraint on the period
(Fig.~\ref{stringlength}).

\begin{figure}
\resizebox{\hsize}{!}{\includegraphics{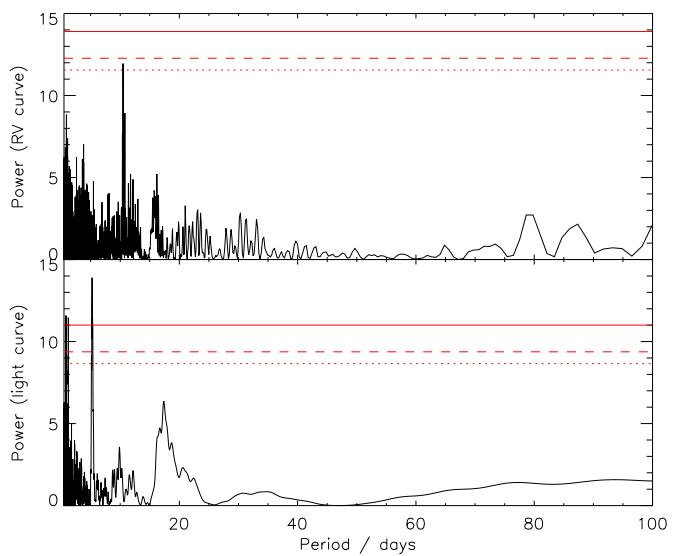}}
\caption{Lomb-Scargle periodograms for F2 WNL RV curve
  (upper panel) and light curve (lower panel), in range 0.5--100~d.
  The red horizontal lines indicate thresholds for false alarm
  probabilities of 0.10, 0.05 and 0.01, from bottom to top, in each
  panel.}
\label{periodogram}
\end{figure}

As a check, a Lomb-Scargle periodogram of the primary RV
data, searching for periods within a much wider range (0.5--100~d),
also finds a maximum peak at 10.483~d.  The same search for the light
curve data finds the strongest peak at around 5.25~d
i.e. approximately half the period (Fig.~\ref{periodogram}).  This is
to be expected for eclipsing binary light curves with eclipses of
similar depth, since the Lomb-Scargle method is equivalent to fitting
sinusoidal functions to the time series, and binary curves are far
more sinusoidal in shape when both eclipses are stacked on top of each
other.  We therefore take 10.483$\pm$0.002~d as the period for the
rest of the analysis.

We use the primary RV curve to determine which of the
light curve minima to take as phase zero, and take BJD 2454679.553 as
our reference zero point.  All phases given in Tables~\ref{rvtable}
and \ref{lctable} are assigned on the basis of this BJD$_0$ and the
period found as described above.

\section{Modelling}

A binary model was developed using the PHOEBE interface
\citep{prsa2005} to the Wilson and Devinney eclipsing binary modelling
code \citep{wilson1971}.  Two approaches were trialled: an
unconstrained binary model, and one using constraints for a W~UMa-type
contact binary (i.e. both components share an effective temperature
$T_{eff}$ and Kopal potential $\Omega$) as the best available
approximations to an assumed close, colliding-wind binary.  Of course,
it may be that the extended, semi-transparent atmosphere of the
Wolf-Rayet component is what is actually in contact with the
secondary, rather than its opaque core, and a more specialized
approach would be required for optimal modelling (as in
e.g. \citealp{perrier2009}).  However, here, given our rather limited
data, we attempt merely to characterize the system in fairly broad
terms.

The effective temperature for the two components ($\tau$=2/3) was
taken as 34\,000~K following spectral modelling (described below), or
as $T_{eff,1}$=34\,100~K and $T_{eff,2}$=33\,800~K for the
unconstrained model.  Other starting assumptions were: bolometric
albedo $\alpha$=0.5 and gravity darkening exponent $\beta$=1.0 for
radiative atmospheres \citep{hilditch2001}; a logarithmic limb
darkening law; and orbital period and zero phase as given above.  The
eccentricity and argument of periastron were first determined by
matching the synthetic curves to the well-defined observed light curve
eclipse phases and the slight asymmetry of the observed primary RV
curve; then the RV curves were used to estimate systemic velocity
$\gamma_0$, orbital separation $a$~sin~$i$ (with orbital inclination
$i$ initially set to 90$^{\circ}$) and mass ratio $q$
($\frac{M_2}{M_1}$, where $M_1$ is the assumed WNL primary).  Finally,
the light curve shape was used to refine the orbital inclination $i$
and Kopal potentials $\Omega_{1,2}$, with $a$ being simultaneously
adjusted to keep $a$~sin~$i$ constant.

\begin{figure}
\resizebox{\hsize}{!}{\includegraphics{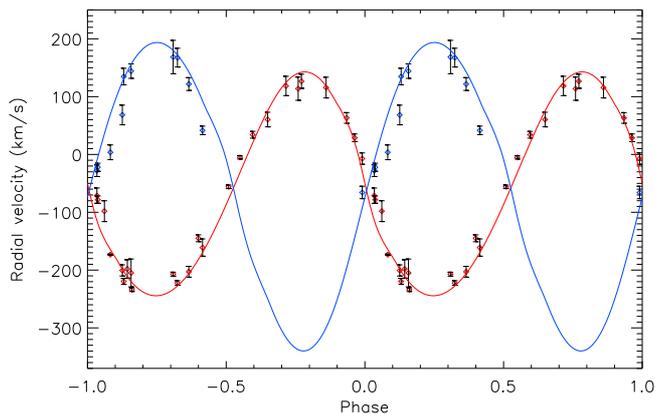}}
\caption{RV curves for F2 (primary in red, secondary in blue), with
  best-fit (unconstrained) PHOEBE models for both components
  overplotted.}
\label{fitrvplot}
\end{figure}

\begin{figure}
\resizebox{\hsize}{!}{\includegraphics{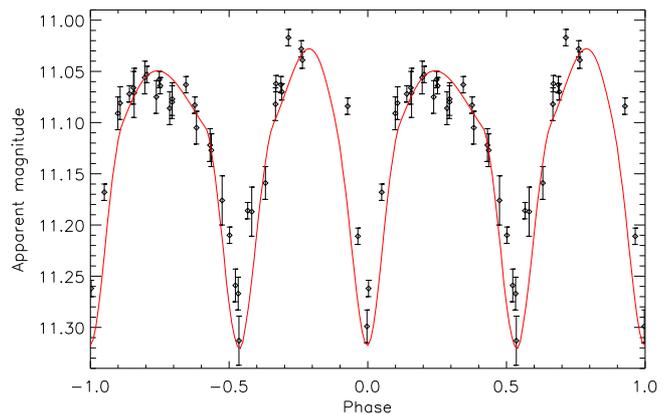}}
\caption{Light curve for F2 with best-fit (unconstrained) PHOEBE model
  overplotted in red.}
\label{fitlcplot}
\end{figure}

\begin{table}
  \caption{Best-fit system and stellar component parameters for F2
    from (unconstrained) binary modelling.  Uncertainties on masses
    are dominated by the error in $a$.}
\label{fittable}
\centering
\begin{tabular}{l | l l}
\hline\hline & Primary & Secondary \\
\hline Semi-major axis $a$ (R$_{\sun}$) & \multicolumn{2}{|c}{105$\pm$5} \\
Mass ratio $q$ & \multicolumn{2}{|c}{0.73$\pm$0.07} \\
Angle of incl. $i$ (\degr) & \multicolumn{2}{|c}{67$\pm$1} \\
Eccentricity $e$ & \multicolumn{2}{|c}{0.075$\pm$0.015} \\
Periastron argument $\omega$ & \multicolumn{2}{|c}{$\frac{\pi}{4}\pm\frac{\pi}{16}$} \\
Periastron phase & \multicolumn{2}{|c}{0.785$\pm$0.005} \\
Systemic velocity $\gamma_0$ (km~s\textsuperscript{-1}) & \multicolumn{2}{|c}{$-$60$\pm$8} \\
Velocity semi-amplitude $K$ (km~s\textsuperscript{-1}) & 203$\pm$9 & 254$\pm$9 \\
Kopal potential $\Omega$ & 3.51$\pm$0.13 & 3.51$\pm$0.13 \\
Surface gravity (log~$g$) & 3.15$\pm$0.02 & 3.14$\pm$0.02 \\
Mass (M$_{\sun}$) & 82$\pm$12 & 60$\pm$8 \\
Radius (R$_{\sun}$) & 40.1$\pm$2.5 & 34.6$\pm$2.0 \\
Luminosity (log~$\frac{L}{L_{\sun}}$) & 6.27$\pm$0.05 & 6.13$\pm$0.05 \\
\hline
\end{tabular}
\end{table}

Owing to the relatively small number of data points of different
origins and quality, and the differing usefulness of points at
different phases in constraining the fits (e.g. during eclipse for the
light curve, or at quadrature for the RV curves), manual adjustment
was found to give a more convincing result than automated convergence
of parameters (Figs.~\ref{fitrvplot} and \ref{fitlcplot}).
Uncertainties in input and output parameters were estimated by varying
input parameters from the best fit while maintaining a plausible match
of the synthetic curves to the observed light and RV curves.  The
resulting best-fit parameters from the unconstrained binary model are
given in Table~\ref{fittable}.

An almost identical solution was obtained using the constraints for a
contact binary, differing in a few details ultimately because of the
shared temperature assumption; crucially, we may note that under both
approaches the best fit requires approximately equal values of
$\Omega_{1,2}$, just outside the Roche limit at the inner Lagrangian
point ($\Omega(L_1)$) i.e. the components of F2 are almost in contact.
No difference in the final component masses was produced by the two
modelling approaches; the radii and hence luminosities of the
components are very slightly smaller under the unconstrained model.

We were then able to disentangle the two components using KOREL
\citep{hadrava2012}, implemented on the Virtual
Observatory\footnote{https://stelweb.asu.cas.cz/vo-korel}.  Our
derived period and epoch of periastron, together with our modelled
mass ratio, semi-amplitude of the primary RV curve, and
eccentricity were fixed as constraints on the disentangling of the
SINFONI spectra, excluding the lower S/N spectrum at phase 0.034,
after their division into six wavelength sections to prevent
undulation of the continuum.  (Allowing these parameters to vary did
not result in a convergent solution.)

\begin{figure*}
\centering
\includegraphics[width=17cm]{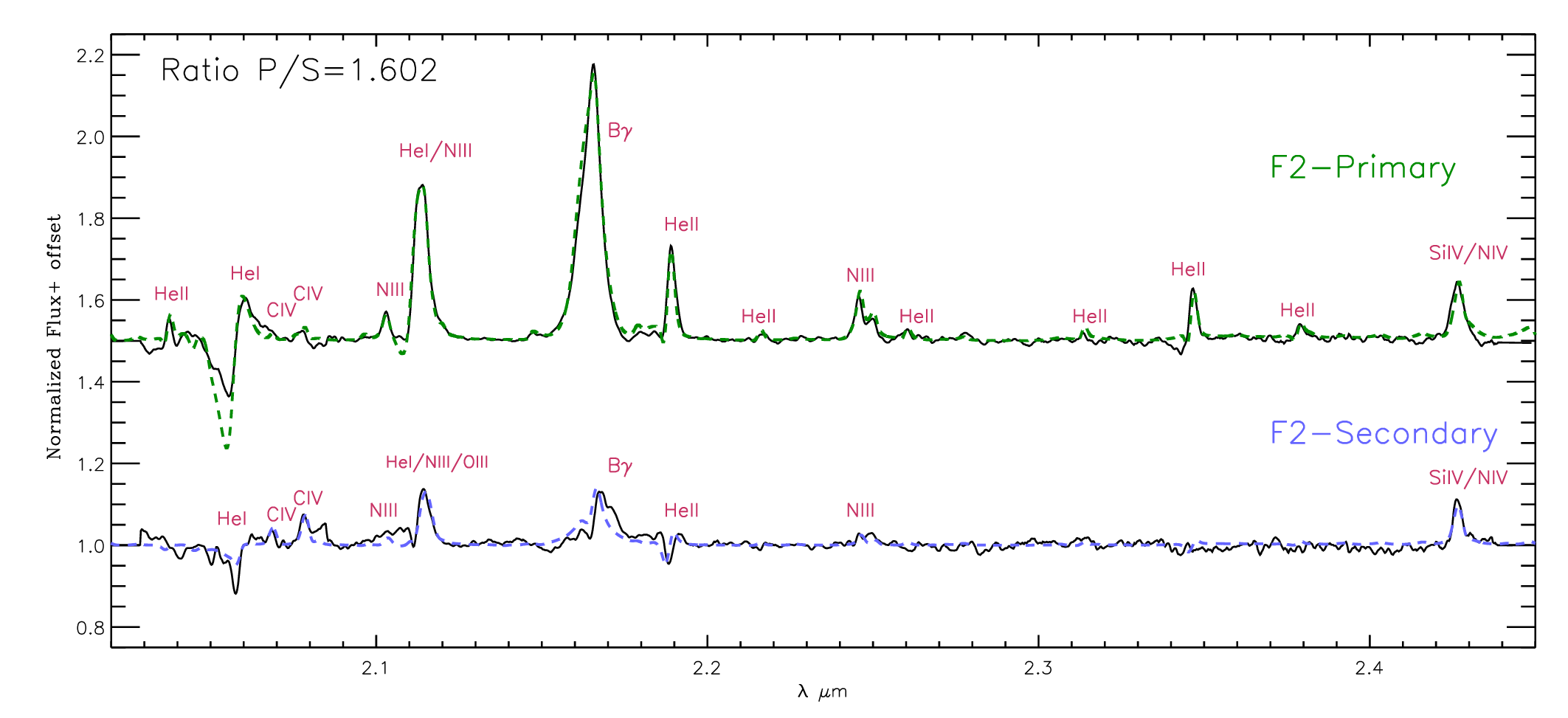}
\caption{Disentangled spectra for F2 primary and secondary components
  (black lines) with best-fit model spectra overplotted (green and blue
  dashed lines).}
\label{disfits}
\end{figure*}

The resulting spectra (rescaled using our preferred derived flux ratio
between components, obtained as described below) are shown in
Fig.~\ref{disfits}.  The primary component appears -- as expected -- to
be a WNL star (Paper I classified it as WN8--9h), while the secondary
shows features of an O hypergiant e.g. Br~$\gamma$ emission;
\ion{He}{I} and \ion{He}{II} absorption.  Paper I placed the secondary
in the context of seven other extreme O super/hypergiants in the
Arches, and gave it a provisional classification as O5--6 Ia$^+$; F17
appears to provide the closest spectral match.  This successful
disentangling, producing a pair of spectra closely resembling other
massive cluster members, gives support to the validity of our binary
modelling and the pair of RV curves on which the model
depends.

Further confirmation was sought through a series of spectral fitting
tests, using high S/N SINFONI spectra at four phases: near the two
eclipses and near the two quadratures.  Levenberg-Marquardt fits were
carried out for more than 400000 combinations of possible primary and
secondary spectra, drawn from two grids of CMFGEN models
\citep{hillier1998,hillier1999}, self-computed to encompass the
parameter domains of interest.  A WNL grid for the primary covered
$T_{eff}$ from 30\,000 to 39\,000~K, log~g from 3.2 to 3.7, wind
density parameter log~$Q$ from $-$11.8 to $-$10.5, He/H from 0.2 to
1.0 (number), and clumping factor $f_{cl}$ from 1 to 0.03.  The
secondary grid consisted of the primary grid (to investigate the
possibility of two WNLs) combined with an O star grid covering dwarf,
giant and supergiant domains; this led to a range in log~g from 3.1 to
4.2 and in log~$Q$ from $-$12.3 to $-$10.5.  Fairly complete model
atoms for C, N, O, Si and S were assumed, to account for the infrared
transitions among high-lying levels present in the K band.  In the
fitting tests, the flux ratio between the two components ($F_{12}$)
and their RVs could also be estimated, and specific
lines could be given more weight in the fitting (e.g. the CIV line at
2.079~$\mu$m, of particular relevance for the secondary component).

\begin{figure*}
\centering
\includegraphics[width=17cm]{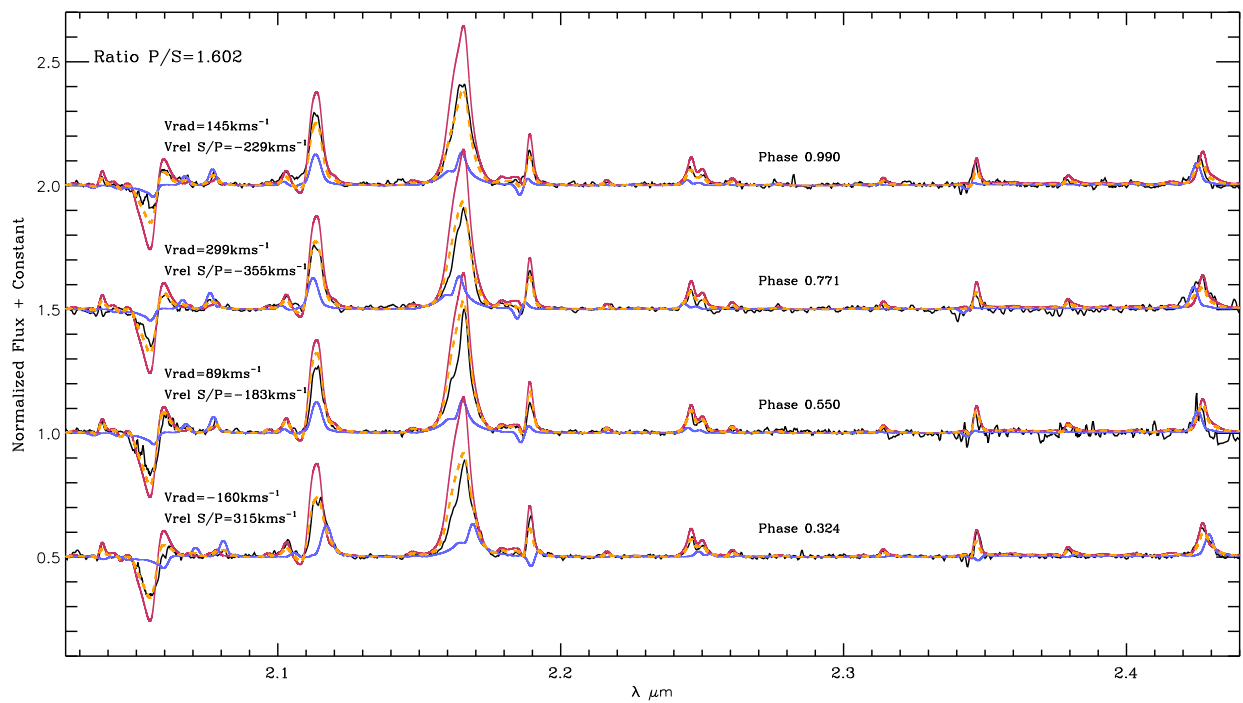}
\caption{Best-fit model spectra for primary (pink), secondary (blue)
  and combined (yellow) components of F2, matched to observed spectra
  (black) at four phases.  Relative velocities of the primary and
  secondary were a free parameter under this modelling approach.}
\label{freefit}
\end{figure*}

\begin{figure*}
\centering
\includegraphics[width=17cm]{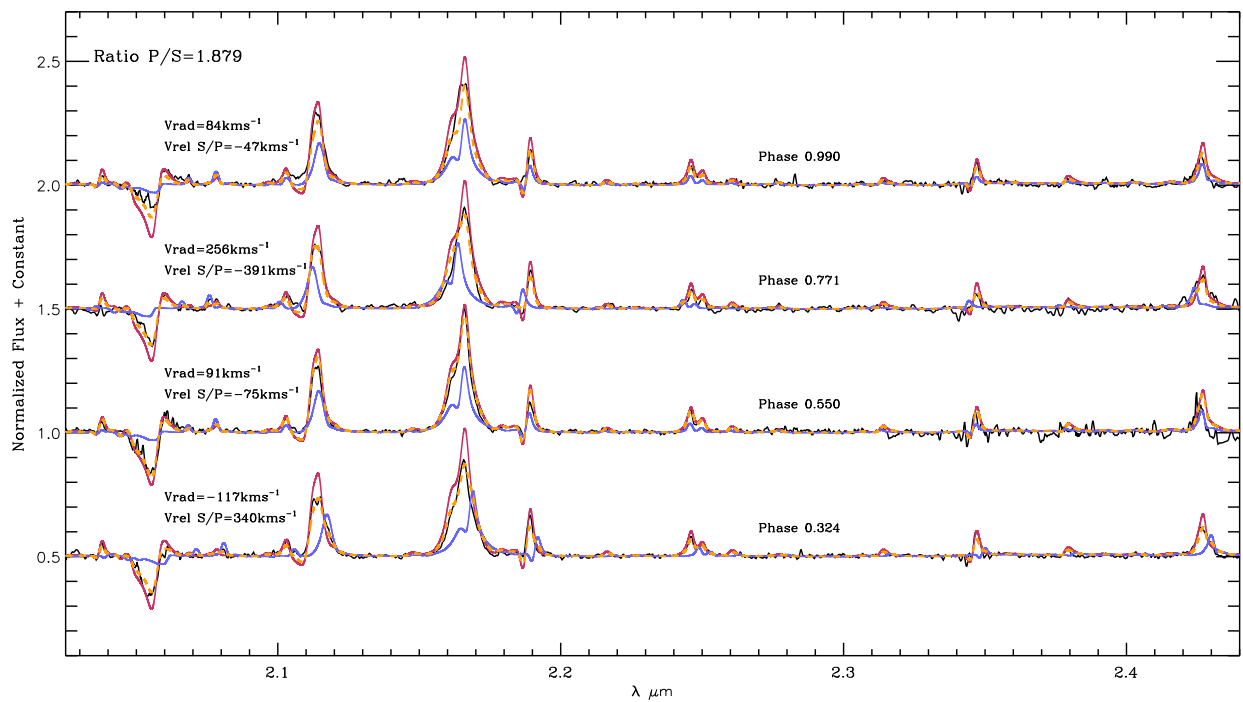}
\caption{Best-fit model spectra for F2, as in Fig.~\ref{freefit}, but
  with relative velocities of the primary and secondary fixed to
  values found from binary modelling.}
\label{fixfit}
\end{figure*}

By fitting all four phases simultaneously, and leaving $F_{12}$ and
the RVs as free parameters, good matches were achieved
to all four observed spectra, using a flux ratio of 1.60,
$T_{eff}\sim$34000~K and log~g$\sim$3.3 for both components
(Fig.~\ref{freefit}).  (The blend of \ion{N}{IV} with \ion{Si}{IV} at
2.428~$\mu$m limits the sensitivity of this otherwise useful line to
log~g for the primary; \ion{He}{II} lines in the secondary are
consistent with log~g between $\sim$3.1 and 3.3 however.)  We should
note that the relative velocity of the two components is
ill-constrained near secondary eclipse where the secondary's weaker
features are largely obscured.  An alternative approach was to fix the
relative velocities of the two components (using the observed radial
velocities and their best-fit modelled curves in Fig.~\ref{fitrvplot})
and leave the flux ratio free; this yielded nearly identical spectral
models to the first method (though with slightly poorer fits to the
combined spectra), and a rather higher $F_{12}$ of 1.88
(Fig.~\ref{fixfit}).

\begin{table*}
\caption{Best-fit stellar component parameters for F2 from CMFGEN
  model fitting to disentangled spectra and from SED fitting.}
\label{cmfgenparams}
\centering
\begin{tabular}{l | l l}
\hline\hline
 & Primary & Secondary \\
\hline
Temperature $T_{eff}$ (K) & 34100$^{+2000}_{-1000}$ & 33800$^{+2000}_{-1000}$ \\
He/H (number) & 0.85$^{+0.15}_{-0.35}$ & 0.25$^{+0.15}_{-0.15}$ \\
Clumping factor\tablefootmark{a} $f_{cl}$ & 0.04\tablefootmark{b} & 0.08\tablefootmark{b} \\
N (mass fraction) & 0.03$\pm$0.15~dex & 0.008$\pm$0.2~dex \\
C (mass fraction) & 0.00015$\pm$0.2~dex & 0.0015$\pm$0.15~dex \\
Si (mass fraction) & 0.0017$\pm$0.2~dex & 0.0014$\pm$0.2~dex \\
Terminal velocity $v_{\infty}$ (km~s\textsuperscript{-1}) & 1325$\pm$150 &
2200 (assumed) \\
Velocity field shape parameter\tablefootmark{a} $\beta$ & 1.15 & 1.15 \\
\hline
Moneti law: & & \\
Luminosity (log~$\frac{L}{L_{\sun}}$) & 6.445 & 6.305 \\
Radius (R$_{\sun}$) & 47 & 41.4 \\
Mass loss $\dot{M}$ (M$_{\sun}$~yr\textsuperscript{-1})) &
3.65$\times$10$^{-5}$ & 1.43$\times$10$^{-5}$ \\
\hline
$\alpha$=2.29: & & \\
Luminosity (log~$\frac{L}{L_{\sun}}$) & 5.648 & 5.508 \\
Radius (R$_{\sun}$) & 19 & 16.5 \\
Mass loss $\dot{M}$ (M$_{\sun}$~yr\textsuperscript{-1})) & 9.23$\times$10$^{-6}$ & 3.61$\times$10$^{-6}$ \\
\hline
\end{tabular}
\tablefoot{
\tablefoottext{a}{More details on the $f_{cl}$ and $\beta$ parameters
  used here are given in \citet{najarro2009}. A 2-$\beta$ law is used
  with $\beta$=1.15 in the inner wind and $\beta$=2.2 in the outer
  wind.}
\tablefoottext{b}{Clumping values between 0.1 and 0.04 are possible
  (with corresponding rescaling).}
}
\end{table*}

Having obtained best-fit model spectra very similar to the disentangled
spectra, a final fit was carried out to each disentangled component
separately, assuming a flux ratio of 1.60.  The resulting fits are
shown in Fig.~\ref{disfits} and provide good matches almost everywhere
except around the He I 2.058~$\mu$m line, where strong and
variable telluric absorption is almost impossible to remove cleanly.
The final parameters of the component models are given in
Table~\ref{cmfgenparams}.

\begin{table}
\caption{Phase-corrected magnitudes used for SED fitting.}
\label{seddata}
\centering
\begin{tabular}{l l l}
\hline\hline
Filter & Magnitude & Uncertainty \\
\hline
F110W & 17.97 & 0.07 \\
F160W & 13.52 & 0.01 \\
F205W & 11.31 & 0.003 \\
F127M & 16.891 & 0.011 \\
F139M & 15.445 & 0.008 \\
F153M & 14.057 & 0.009 \\
F190N & 11.830 & 0.004 \\
\hline
\end{tabular}
\end{table}

To provide an independent estimate of the luminosities of the
components of F2, SED modelling of the system during secondary eclipse
was carried out by combining the primary and secondary best-fit
models, assuming a particular $K$-band flux ratio between the
components, and using the difference in system magnitude relative to
the out-of-eclipse magnitude of 0.264 (obtained from the fitted light
curve).  The observed magnitudes used, adjusted for phase, are given
in Table~\ref{seddata}.  Given the uncertainty over the appropriate
extinction prescription to use, a variety of extinction laws were
explored, as described for the wider cluster in Paper I.  The two main
types were $\alpha$-laws
(i.e. $A_\lambda$=$A_{k0}(\frac{\lambda_{k0}}{\lambda})^\alpha$, where
$A_{k0}$ is the monochromatic value at 2.159~$\mu$m = $\lambda_{0}$ in
2MASS $K_s$), and Moneti's law, where $R_V$ is assumed to be 3.1, and
$A_{k0}$ is derived.  (Further tests were performed assuming fixed
$\alpha$ values from literature or by using a two-$\alpha$ approach.)

\begin{figure*}
\centering
\includegraphics[width=17cm]{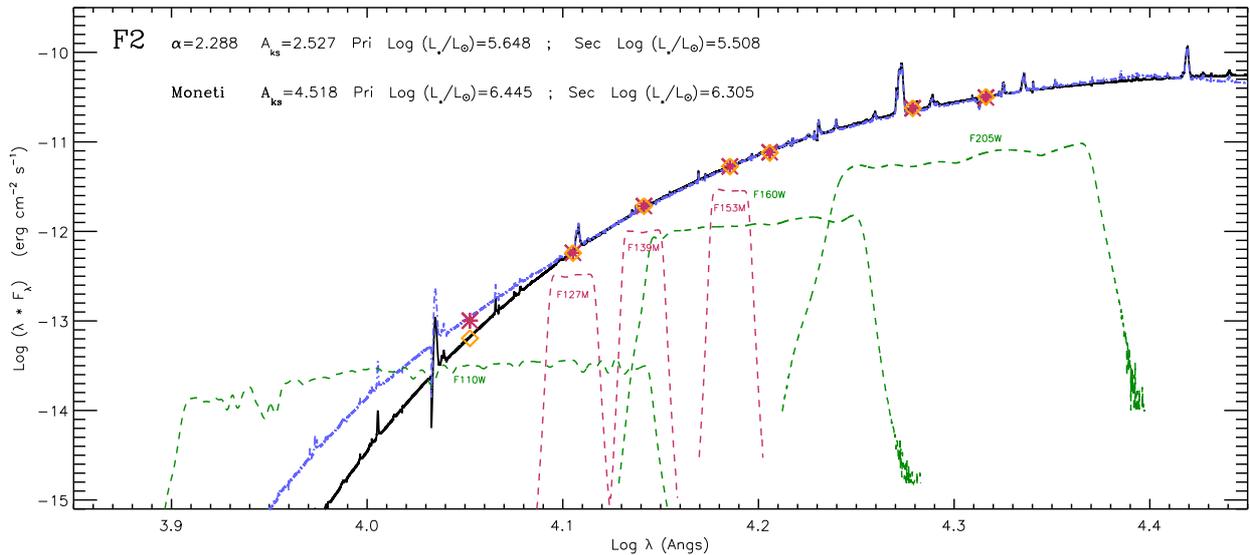}
\caption{Best-fit SEDs for F2 using different extinction
  prescriptions.  In black, the preferred primary and secondary models
  are combined, with final luminosities of
  log~$\frac{L}{L_{\sun}}$=5.648 and 5.508 respectively, and are
  reddened with $\alpha$=2.288 and $A_{ks}$=2.527.  The blue line uses
  luminosities of log~$\frac{L}{L_{\sun}}$=6.445 and 6.305, and
  reddening follows Moneti's law with $A_{ks}$=4.518.  Filter curves
  for the filters used for the fit are shown in green (broadband) and
  pink (narrowband), and symbols are plotted for each magnitude
  measurement to show the goodness of fit: yellow diamonds for the
  $\alpha$-model and pink stars for the Moneti model.  (The filter
  curve for F190N is not shown as it was not used for fitting,
  although it matches the curves very well.)  The $x$-axis position of
  each symbol corresponds to the classical $\lambda_0$ of the filter
  at which the zero-point flux is defined.  The $y$-axis position
  coincides with its corresponding model curve if the observed
  magnitude matches the magnitude of the reddened model.}
\label{sedfit}
\end{figure*}

Fig.~\ref{sedfit} and Table~\ref{cmfgenparams} show the best-fit
results for the two main extinction approaches, assuming
$F_{12}$=1.60.  It is apparent that both extinction prescriptions are
capable of fitting the observed magnitudes well, but they lead to
differences in $A_{k0}$ of almost two magnitudes for such a
heavily-reddened object as F2, which translates into 0.8 dex in log
$L$.  The plot also shows the degeneracy in the extinction in the
4.1--4.4 log $\lambda$ region, which would require medium-width
filters at 1.0 and 1.1 $\mu$m to break.  However, only the Moneti
approach yields luminosities roughly compatible with our binary
modelling results (Table~\ref{fittable}), even allowing for
uncertainties in the SED-modelled log~$\frac{L}{L_{\sun}}$ of around
+0.08, -0.04 dex.  Stellar radii below 20~$R_{\sun}$ would also imply
implausibly low masses in the context of the Wolf-Rayet members of the
Arches.

\begin{figure}
\resizebox{\hsize}{!}{\includegraphics{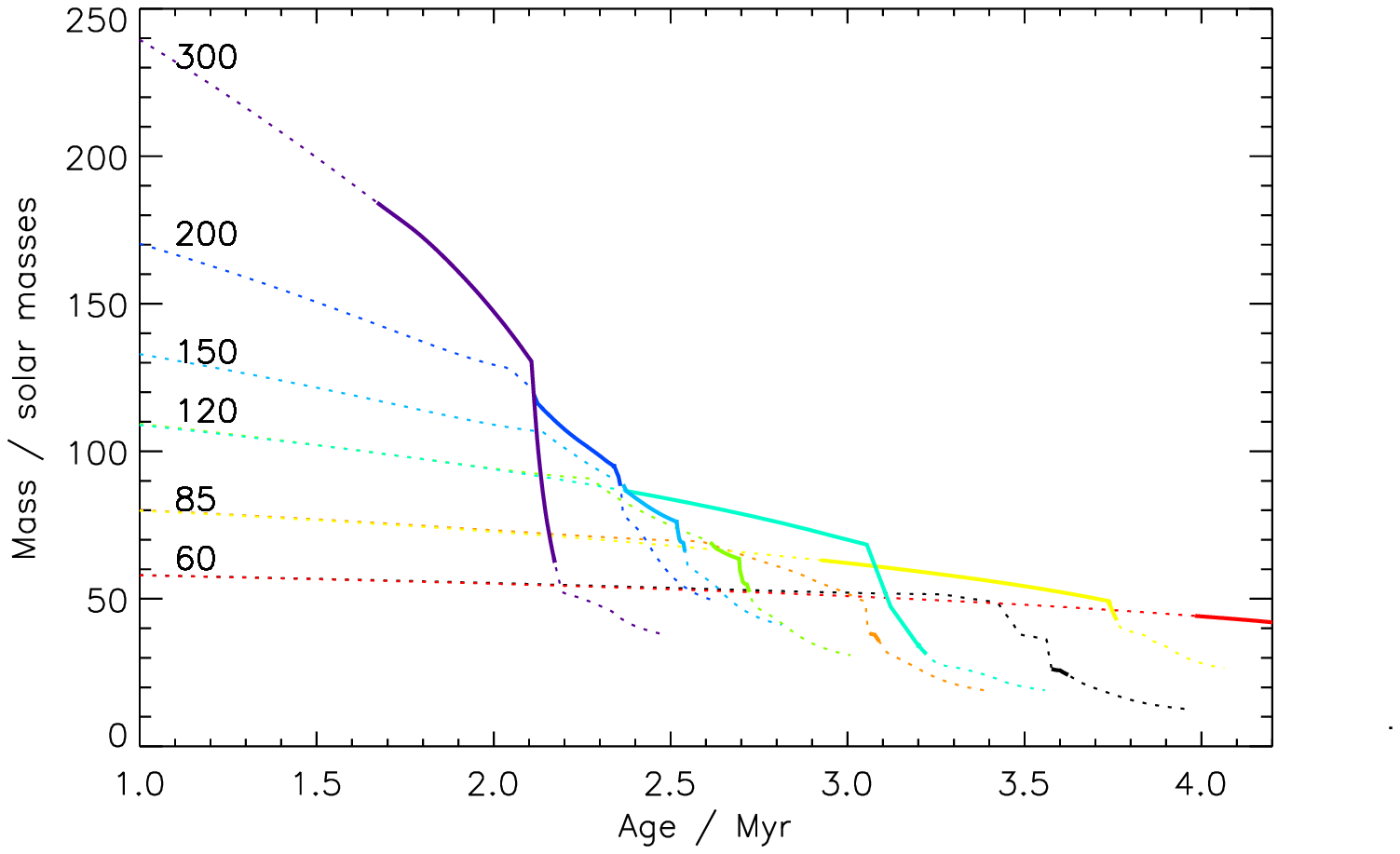}}
\caption{Mass over time for candidate primary component of F2, from
  latest Geneva evolutionary models with $Z$=0.014.  Initial masses
  are shown (in solar masses) next to each track on the left; for the
  60, 85 and 120~M$_{\sun}$ tracks, both rotating and non-rotating
  models are shown (the rotating models are in red, yellow and cyan
  respectively, and have greater duration than the corresponding
  non-rotating versions in black, orange and green); for higher
  initial masses, only non-rotating models are available.  Each track
  follows the life of a star to the end of the C burning stage (dotted
  lines); WNL phases, as defined in \citet{georgy2012}, are shown in
  thicker solid lines.}
\label{masstimeplot}
\end{figure}

Given the substantial mass-loss expected for a Wolf-Rayet star through
stellar wind (directly supported in the case of F2 by the P~Cygni
profiles of some lines, and by model atmosphere analysis), it is
useful to compare our model results with the expected mass evolution
of such a star.  Fig.~\ref{masstimeplot} shows Geneva evolutionary
tracks \citep{ekstrom2012} for stars with initial masses
60--300~M$_{\sun}$, including rotating models where available.
Although these models do not include binary interactions, they should
give some indication of the maximum mass loss expected for the
primary, and thus its minimum plausible current mass.  Using an age
for the Arches of 2.5$\pm$0.5~Myr \citep{figer2002,martins2008}, we
can see that the most massive stars would already have completed their
WNL stage, while the least massive would not yet have reached
it\footnote{However, following \citet{martins2017}, caution is
  required in using the definition of a Wolf-Rayet/WNL phase in an
  evolutionary model, based on abundances, to evaluate a
  spectroscopically-defined WNL; the two will not necessarily be
  consistent.}.  With the age of 3.5$\pm$0.7~Myr proposed by
\citet{schneider2014}, the only plausible candidates would be the
rotating 60, 85 and 120~M$_{\sun,init}$ tracks, or (very briefly) the
non-rotating 60 and 85~M$_{\sun,init}$ tracks.  Combining these age
constraints would allow a current mass between $\sim$150 and
$\sim$25~M$_{\sun}$; if we prefer rotating models for a binary
configuration, the lowest possible current mass for the primary would
be $\sim$40~M$_{\sun}$.

\begin{figure}
\resizebox{\hsize}{!}{\includegraphics{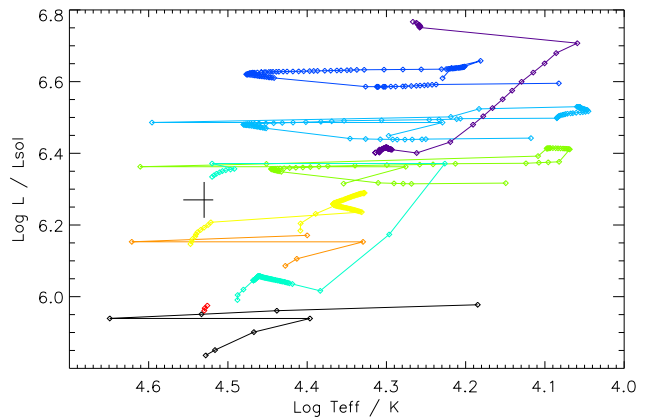}}
\caption{HR diagram of candidate Geneva models for F2 primary star.
  Partial evolutionary tracks are shown, for the WNL stage at ages
  2--4.2~Myr.  Each track has the same colour as in
  Fig.~\ref{masstimeplot}.  The cross indicates the luminosity and
  temperature (with uncertainties) of the primary of F2 from our
  binary and spectral models respectively.}
\label{hrplot}
\end{figure}

We may also look at the temperature and luminosity of F2's primary
according to the model.  Fig.~\ref{hrplot} shows an HR diagram for the
WNL phases of the same evolutionary tracks, where these overlap with
the expected age range for the Arches (2.0--4.2~Myr).  Also shown here
is the location of the modelled F2 primary, using the luminosity found
from binary modelling.  We may observe that it lies close to the
start of the rotating 120~M$_{\sun,init}$ track (in cyan).  Referring
back again to Fig.~\ref{masstimeplot}, and matching its modelled mass
of 82$\pm$12~M$_{\sun}$ with this evolutionary track, an age of
2.6$^{+0.4}_{-0.2}$~Myr is supported.  (If the slightly higher
luminosity found from SED fitting, using the Moneti law, is preferred,
the 150~M$_{\sun,init}$ track would be supported instead, suggesting
an age closer to 2.5~Myr.)

At this age, the secondary's modelled current mass of
60$\pm$8~M$_{\sun}$ would place it in the region of the 60 and
85~M$_{\sun,init}$ tracks, prior to the WNL stage or possibly just at
the start of it (using the rotating 85~M$_{\sun,init}$ track).
However, it should be noted that simulations in \citet{groh2014} were
unable to generate an O hypergiant phase from a 60~M$_{\sun,init}$
star; this may indicate that F2's secondary was originally more
massive.  $R_2$ is perhaps rather large and hence
log~$\frac{L}{L_{\sun}}$ rather high for an O hypergiant secondary:
comparable single stars studied in \citet{martins2008} (F10 and F15)
are modelled as having radii $\sim$30~R$_{\sun}$.  However, an
inflated secondary radius is expected in lower-mass contact binaries
due to the shared envelope \citep{lohr2015}, and a comparable
mechanism may be in operation here.

\section{Discussion}

The most plausible model for F2, based on available data, appears to
be a near-contact eclipsing SB2 binary containing a
$\sim$80~M$_{\sun}$ WN8--9h component with a strong stellar wind, and
a luminous $\sim$60~M$_{\sun}$ O-type secondary.  The orbit seems to
be not yet quite circularized, but some interaction is probable in the
form of colliding winds, which may explain the behaviour of the
\ion{C}{IV} line at 2.079~$\mu$m around and after the secondary
eclipse.  Such colliding winds are supported by detections near the
location of F2 of X-ray source A6 \citep{wang2006}, and radio source
AR10 \citep{lang2005} with log \.{M} of
$-$4.72~M$_{\sun}$~yr\textsuperscript{-1} (where we found log \.{M} of
$-$4.29~M$_{\sun}$~yr\textsuperscript{-1}); non-detection of AR10 in
\citet{lang2001} may also indicate radio variability.

\begin{table*}
\caption{Published parameters for comparable binaries with dynamical
  mass estimates, containing WNL components.}
\label{wnltable}
\centering
\begin{tabular}{l l l l l l}
\hline\hline
\noalign{\smallskip} 
Name & Spectral types & $P_{orb}$ & $M_1+M_2$ & Age & References \\
 & & (d) & (M$_{\sun}$) & (Myr) & \\
\hline
F2 & WN8--9h + O5--6~Ia$^+$ & 10.5 & 82$\pm$12 + 60$\pm$8 &
$\sim$2.6 & (This work) \\
WR21a & O3/WN5ha + O3Vz((f*)) & 31.7 & 104$\pm10$ + 58$\pm4$ & 1--2 & (1,2,3) \\
WR22 & WN7+abs + O9 & 80.3 & 55$\pm7$ + 20$\pm2$ & ?2.2 & (4,5,6) \\
WR20a & WN6ha + WN6ha & 3.7 & 83$\pm5$ + 82$\pm5$ & 1--2 & (7,8,9) \\
NGC~3603-A1 & WN6ha + (?)WN6ha & 3.8 & 116$\pm31$ + 89$\pm16$ &
$\sim$1 & (10,11) \\
R145 & WN6h + O3.5~If*/WN7 & 159 & $\sim$80 + $\sim$80 & $\sim$2.2 & (12,13) \\
R144 & WN5--6h + WN6--7h & $<370$ & ?90 + ?120 & ?2 & (14) \\
WR25 & WN + O & 208 & ?75$\pm7$ + ?27$\pm3$ & $\sim$2.5 & (15,16) \\
WR29 & O + WN7h & 3.2 & 53$\pm4$ + 42$\pm4$ & -- & (17,18) \\
WR77o & O + WN7o & 3.5 & 43$\pm7$ + 16$\pm3$ & 4.5--5 & (19,20,21) \\
\hline
\end{tabular}
\tablebib{(1)~\citet{benaglia2005}; (2) \citet{niemela2008}; (3)
  \citet{tramper2016}; (4) \citet{rauw1996}; (5)
  \citet{schweickhardt1999}; (6) \citet{grafener2008}; (7)
  \citet{rauw2004}; (8) \citet{bonanos2004}; (9) \citet{naze2008}; (10)
  \citet{moffat2004}; (11) \citet{schnurr2008}; (12)
  \citet{schnurr2009}; (13) \citet{shenar2017}; (14)
  \citet{sana2013b}; (15) \citet{gamen2008}; (16) \citet{hur2012};
  (17) \citet{niemela2000};
  (18) \citet{gamen2009}; (19) \citet{negueruela2005}; (20)
  \citet{crowther2006}; (21) \citet{koumpia2012}.}
\end{table*}

F2 thus seems comparable with a handful of very massive binaries
containing WNL stars as primaries.  WR21a
\citep{benaglia2005,niemela2008,tramper2016} is particularly similar:
an X-ray bright, colliding-wind, eccentric spectroscopic O3/WN5ha+O3
binary with $P$=31.7~d, and with absolute masses for its components of
$\sim$104 and 58~M$_{\sun}$; the secondary's lines were only detected
from disentangled high-resolution spectra.  WR22
\citep{rauw1996,schweickhardt1999,grafener2008} is a much longer
period binary containing rather less massive WN7+abs and O9
components, of which \citeauthor{schweickhardt1999} noted ``the
absorptions from the companion are extremely weak and they can only be
detected in spectra with a very high signal to noise ratio''.  An even
longer period and highly eccentric binary, R145
\citep{schnurr2009,shenar2017}, has recently been disentangled to
reveal WN6h and O3.5If*/WN7 components, with primary and secondary
masses $\sim$55~M$_{\sun}$ (from orbital and polarimetric analysis),
or nearer $\sim$80~M$_{\sun}$ (using quasi-homogeneous evolution
tracks).  WR20a \citep{rauw2004,bonanos2004,naze2008}, also a
colliding-wind binary, has nearly equal-mass WN6ha components (82 and
83~M$_{\sun}$) and is a clear SB2 with a much shorter period than F2
(3.7~d), but its deeply-eclipsing light curve shows a strong
resemblance to F2, and was also modelled as a contact or near-contact
system.  Possibly more massive than F2 is the $P$=3.8~d system
NGC~3603-A1 \citep{moffat2004,schnurr2008} containing a WN6ha and a
secondary of similar spectral type, although the RVs for its secondary
were difficult to determine due to colliding wind features and
blending of the extremely broad lines used, resulting in 20--30\%
uncertainties on masses.  Even more challenging is R144
\citep{sana2013b}, which appears to be an SB2 with very high
luminosity (log~$\frac{L}{L_{\sun}}\sim$6.8), but for which no clear
period has yet been determined.

In contrast, WR29 \citep{niemela2000,gamen2009} is a WN7h+O eclipsing
binary in which the O star is found to be more massive than the
Wolf-Rayet (53 and 42~M$_{\sun}$).  Its light curve appears very
different from that of F2, with primary and secondary eclipses and
maxima of different depths and heights; also, a number of absorption
lines from the O star were readily observed in its spectra,
superimposed on the Wolf-Rayet's emission lines.  A less massive
system of this type appears to be WR77o
\citep{negueruela2005,crowther2006,koumpia2012}, with a
$\sim$43~M$_{\sun}$ O-type primary and a $\sim$16~M$_{\sun}$ WN7o
(hydrogen-depleted) secondary; the O star's absorption lines were not
detected there in spite of its greater temperature and radius, but the
system is still not a good comparator for the highly luminous and
hydrogen-rich F2.  Details for these and other WNL binary members are
given in Table~\ref{wnltable}.

It is notable that the only Galactic binaries on that list which are
currently more massive than F2 (WR21a, WR20a and NGC~3603-A1) are
considerably younger (1--2~Myr), and so may be expected to be closer
to their initial masses.  F2, with an age around 2.6~Myr, and likely
initial primary mass between 120 and 150~M$_{\sun}$, may have been the
most massive binary in the Galaxy at its formation.

In any case, the binarity of F2 is indisputable given its photometric
and spectroscopic variability with a common period.  Might there be
other binaries in the Arches?  Significant photometric variability was
not observed in any bright Arches targets besides F2.  However,
\citet{wang2006} identified three other bright X-ray sources in the
cluster: F6, F7 and F9, and suggested these were produced by
``colliding stellar winds in massive star close binaries''; of these,
F6 and F7 were also detected as bright radio-variable sources in
\citet{lang2001,lang2005} (F9 was not included in their survey).  Our
preliminary spectroscopic results for these three objects (see Paper
III) do provide some support for significant RV
variations, though with far lower amplitudes of variability than F2.
More convincingly, the OIf+ hypergiant F15 and the O supergiant F35
exhibit highly significant variability with $\Delta$RV around 75 and
220~km~s\textsuperscript{-1} respectively.  Forthcoming scheduled
observations of these and other Arches targets should allow us to
confirm their spectroscopic variability status, and establish whether
any variability is periodic and plausibly associated with binarity.

The demonstrable binarity of at least one very massive member of the
Arches may present a challenge to \citet{schneider2014}, who argue
that the most massive 9$\pm$3 stars in this cluster are expected to be
rejuvenated products of binary mass transfer and merger.  Such
rejuvenation is proposed as an explanation for the finding of
\citet{martins2008} that ``the most massive stars are slightly younger
than the less massive stars'', which could otherwise be understood as
caused by an extended star formation period.
\citeauthor{schneider2014} determine a cluster age of 3.5$\pm$0.7~Myr
on this basis, which barely overlaps with our estimated age for F2 of
2.6$^{0.4}_{0.2}$~Myr, at least using evolutionary models for single
stars.  Such models may not be fully appropriate for a near-contact
system, in which some interaction is very likely to have occurred
already, at least between the component winds; however, in the current
absence of suitable evolutionary models for such massive binary
components, and given that interaction has apparently not yet been
sufficient to circularise the binary orbit, we regard them as
providing a usable approximation.

It is of course possible that F2 formed more recently than the rest of
the cluster, though this explanation would be less tenable if other
spectroscopically-variable Arches members prove to be binaries as
well.  We may also note that the spectrum of F2's primary is very
similar to those of the other highly-luminous WNL members of the
cluster (Paper I); if the others had been ``rejuvenated'' while this
one had not, we might expect a greater difference.  In Paper I, we
found a smooth progression in spectral morphology from WNLh stars
through OIf+ hypergiants to O supergiants, giants and arguably dwarfs,
supporting coevality; in spite of our survey reaching a magnitude at
which they should have been detectable, no more evolved objects such
as hydrogen-depleted WNE or WC stars were observed, nor transitional
objects such as luminous blue variables, as are seen in the slightly
older Quintuplet cluster \citep{geballe2000}.

\section{Conclusions}

Our spectroscopic survey of the most massive and luminous members of
the Arches cluster revealed one source which exhibits significant and
substantial RV variability: F2, classified as WN8--9h by
\citet{martins2008}.  Further archival spectra indicated that the
velocities vary sinusoidally, with a period of 10.483~d, matching the
light curve period found by \citet{markakis2011} for the same source
in a photometric survey.  Re-reducing the photometry yielded a light
curve typical of a contact or near-contact eclipsing binary, with a
small eccentricity.  Radio and X-ray observations also suggested F2
might be a colliding-wind massive binary.

We have presented preliminary models for this system as an SB2
near-contact eclipsing binary (the secondary RV curve is
only partially determined by a single faint line moving in anti-phase
with the WNL primary).  Combining the RV and light
curves, and using an effective temperature obtained from spectral
fitting, we model the system as having $M_1$=82$\pm$12~M$_{\sun}$,
$M_2$=60$\pm$8~M$_{\sun}$.  The primary's modelled properties are
consistent with a rotating 120~M$_{\sun,init}$ evolutionary track and
an age of 2.6$^{+0.4}_{-0.2}$~Myr, though an initial mass as high as
150~M$_{\sun}$ and age nearer 2.5~Myr is also consistent with our
SED-modelled luminosities.  From spectral disentangling and
independent searches for best-fitting template spectra from a large
grid of CMFGEN models, the secondary is most likely an O hypergiant
similar to others in the Arches (O5--6 Ia$^+$).

F2 therefore appears to be one of a small set of very massive binaries
containing Wolf-Rayet primaries, and is perhaps the most massive
binary in the Galaxy in terms of initial mass, since other systems
with higher dynamical mass estimates are in much younger clusters and
so will have lost less mass.  It is also the first confirmed binary in
the Arches cluster, potentially conflicting with an argument of
\citet{schneider2014} that the most massive members of the Arches are
expected to be the products of binary interaction and merger, and that
the cluster itself is around 3.5~Myr old.  F2 is one of four hard,
bright X-ray sources within the Arches that have been interpreted as
colliding-wind binaries; the confirmation of its binary nature
strongly implies that the other three sources are similar systems.
Moreover, our spectroscopic survey has revealed a number of additional
RV variables, although further observations will be required to
characterise these fully (Paper III).  A synthesis of these findings
hints at a rich binary population within the Arches.

The confirmation of the binary nature of F2 alone has important
implications for the global properties of the Arches, potentially
allowing us to calibrate the mass--luminosity relationship and
consequently determine the cluster's (initial) mass function and
integrated mass.  In turn, these parameters will enable us to place
constraints on the upper mass limit of stars; the apparently
pre-interaction configuration of F2 also has important implications
for the formation of very massive stars, suggesting that they do not
all form via binary mass transfer and/or merger.  Finally,
determination of the frequency of occurrence of such very massive
binaries will provide critical observational constraints for
population synthesis modelling of the progenitors of massive,
coalescing, relativistic binaries and gravitational wave sources.

\begin{acknowledgements}
Based on observations collected at the European Organisation for
Astronomical Research in the Southern Hemisphere under ESO programmes
075.D-0736, 081.D-0480, 087.D-0317, 087.D-0342, 091.D-0187 and
093.D-0306, and based on data obtained from the ESO Science Archive
Facility under request numbers 155505, 155514, 189886, 190865, 283402,
283411, 283415, 283417 and 283512.  This research was supported by the
Science and Technology Facilities Council.  We are grateful to Dong
Hui for assistance with the reduction of archival photometric data for
our spectral energy distribution modelling, and to Fabrice Martins for
the kind provision of his SINFONI reduced data cubes from 2005, as
well as his design of the original SINFONI observing strategy used for
the Arches.
\end{acknowledgements}

\bibliographystyle{aa}
\bibliography{archesrefs}

\end{document}